\documentclass[10pt,twocolumn,letterpaper]{article}

\usepackage[utf8]{inputenc}
\usepackage[T1]{fontenc}
\usepackage[margin=0.75in,top=0.75in,bottom=1in]{geometry}
\usepackage{graphicx}
\usepackage{hyperref}
\usepackage{url}
\usepackage[english]{babel}
\usepackage{adjustbox}
\usepackage{tabularx}
\usepackage{booktabs}
\usepackage{makecell}
\usepackage{subcaption}
\usepackage{amsmath}
\usepackage{amssymb}
\usepackage{titlesec}
\usepackage{abstract}
\usepackage{parskip}

\titleformat{\section}{\normalfont\bfseries\scshape}{\Roman{section}.}{0.5em}{}
\titleformat{\subsection}{\normalfont\itshape}{\Alph{subsection}.}{0.5em}{}
\titlespacing*{\section}{0pt}{8pt}{4pt}
\titlespacing*{\subsection}{0pt}{5pt}{2pt}

\makeatletter
\renewcommand{\@biblabel}[1]{[#1]}
\renewcommand{\thebibliography}[1]{%
  \section*{\textbf{References}}
  \small
  \list{\@biblabel{\@arabic\c@enumiv}}{%
    \settowidth\labelwidth{\@biblabel{#1}}%
    \leftmargin\labelwidth
    \advance\leftmargin\labelsep
    \usecounter{enumiv}%
    \let\p@enumiv\@empty
    \renewcommand\theenumiv{\@arabic\c@enumiv}}%
  \sloppy\clubpenalty4000\widowpenalty4000%
  \sfcode`\.\@m}
\makeatother

\title{\LARGE \textbf{Towards Revised Tempo Indications for Beethoven's Piano and Cello Sonatas: Czerny, Moscheles, Kolisch, and Recorded Practice 1930--2012}}

\author{Dr Ignasi Sole \quad \texttt{ignasiphd@gmail.com} \quad \today}

\begin{document}

\maketitle
\thispagestyle{empty}
\pagestyle{empty}


\begin{abstract}
Historical metronome indications for Beethoven's five piano and cello sonatas (as transmitted by Czerny, Moscheles, and Kolisch), have long been regarded as problematic by performers and scholars alike. This paper presents the first systematic empirical assessment of those indications against a corpus of over one hundred movement-level recordings spanning 1930--2012, encompassing first, second, and third movements across all five sonatas (Op.~5 Nos.~1 and~2; Op.~69; Op.~102 Nos.~1 and~2). The core findings are threefold. First, Czerny's and Moscheles's markings are consistently and substantially exceeded by the entire recording corpus: gaps of 15--39\% are documented across movements, with the largest divergences in slow Adagio movements and the smallest in fast Allegro finales. Second, Kolisch's 1943 markings align considerably more closely with recorded practice than either Czerny's or Moscheles's, a striking result given that Kolisch was reasoning without corpus data. Third, the central Allegro tempo traditions for each movement are stable across eight decades; not because all performers play alike, but because three coexisting slow, mid-range, and fast traditions persist simultaneously, with the mid-range dominant throughout. Building on these findings, this paper proposes a set of revised tempo indications grounded in the statistical modal tempi of the corpus, presented as ranges reflecting the documented spectrum of expert interpretive practice rather than single prescriptive values. These indications are offered not as claims about Beethoven's intentions but as evidence-based reference points for performers and scholars navigating the gap between historical prescription and performable reality.
\end{abstract}


\section{Introduction}

The tempos Beethoven attached to his works have generated debate since the 1810s, when performers first encountered markings that seemed implausibly fast. The debate intensified with the spread of Maelzel's metronome and has never fully resolved. Three interventions define its modern shape. Czerny's and Moscheles's nineteenth-century metronome marks, transmitted in their editions and commentaries, set the standard reference point that most subsequent discussion has either defended or challenged. Kolisch's 1943 study~\cite{c1} argued that the markings are reliable and performers simply too cautious, while proposing alternative indications for many works based on his interpretation of musical character. Noorduin's systematic compilation of tempo marks from Beethoven's students and contemporaries~\cite{c2} provided the most rigorous historical account yet of how editorial interventions shaped received performance traditions, arguing that the slow-tempo tradition in late Beethoven was a construction of nineteenth-century editors rather than a reflection of Beethoven's intentions.

What has been absent from this debate until now is systematic empirical evidence: not argument about what the markings mean or should mean, but data showing what expert performers across successive generations have actually done. This paper supplies that evidence for a specific and previously unstudied repertoire: Beethoven's five sonatas for piano and cello. It analyses over one hundred movement-level recordings spanning 1930--2012, compares the resulting tempo data against the indications of Czerny, Moscheles, and Kolisch, and proposes revised tempo indications grounded in that comparison.

The paper does not claim to resolve what Beethoven intended. It makes a more modest and more actionable claim: that the range of tempos adopted by demonstrably expert performers across eight decades constitutes a meaningful constraint on the space of viable interpretations, and that indications grounded in that range are more useful to performers and scholars than indications grounded in theoretical argument alone.

The paper is structured as follows. Section~II reviews the principal explanatory frameworks for the gap between historical markings and recorded practice. Section~III describes the corpus and methodology. Sections~IV--VI present the tempo, performance time, and portamento findings. Section~VII presents the proposed revised indications. Section~VIII discusses limitations and future directions. Section~IX concludes the paper.


\section{Background}

\subsection{Why the Markings May Be Unreliable}

Several distinct hypotheses have been advanced to explain why Beethoven's metronome markings consistently exceed what performers find viable. They are not mutually exclusive and may all have contributed.

The most widely cited is mechanical: Beethoven's metronome, one of Maelzel's earliest designs, had documented reliability problems. Beethoven's own letters record complaints about the device's ``uneven pulse'' and frequent need for repair~\cite{c3}. Fors\'en, Gray, Lindgren, and Gray argue that a slipped weight or fallen mechanism could distort the indicated tempo by a significant margin~\cite{c4}. Mart\'in-Castro and Ucar propose a more specific mechanism: that Beethoven may have misread the triangular measurement indicator of the metronome, potentially adding approximately 12~BPM to his markings~\cite{c5}. If either hypothesis is correct, the markings would be systematically inflated relative to Beethoven's actual intentions.

The second hypothesis is instrumental. Newman reports that Beethoven considered the Erard piano's heavy action cumbersome for rapid passages~\cite{c6}. Skowroneck adds that the lighter Viennese mechanisms facilitated faster, more fluid execution and were therefore better suited to his original tempo indications~\cite{c7}. Van Oort and Moschos identify tonal-decay differences between fortepianos and modern instruments as a further factor: the fortepiano's rapid decay under leather damping may have necessitated faster tempi to sustain melodic continuity, whereas the modern piano's longer sustain allows a slower tempo without sacrificing the sonic thread~\cite{c7, c8}. On this account, markings calibrated for the fortepiano would feel implausibly fast on a modern instrument not because they are inaccurate but because they reflect a different acoustic reality.

The third hypothesis concerns the nature of Czerny's and Moscheles's markings themselves. Taruskin argues that many nineteenth-century metronome marks reflect aesthetic ideals rather than empirically tested rehearsal speeds~\cite{c9}. Noorduin's detailed genealogy of editorial interventions shows how figures including Schindler and Wagner reshaped performance traditions in ways that diverged significantly from Beethoven's documented preferences~\cite{c2}. On this account, the markings are not even reliable representations of what Czerny and Moscheles heard in Beethoven's own performances but are theoretical constructions calibrated to a particular aesthetic vision.

\subsection{Kolisch's Position}

Kolisch~\cite{c1} occupies an unusual position in this debate. He argued that Beethoven's markings are essentially reliable and that the received tradition of slow performance is a post-Beethoven construction. He proposed alternative markings for many works (including the piano and cello sonatas) based on his interpretation of their musical character and his own performance experience. His markings are generally more moderate than Czerny's and Moscheles's, sitting between the historical prescriptions and what performers actually do. This paper's empirical finding (that Kolisch's markings align more closely with recorded practice than either Czerny's or Moscheles's) is therefore both a confirmation and a complication of his position: a confirmation because his intuitions tracked something real, a complication because he arrived at those markings by a different route than corpus analysis would suggest.

\subsection{The Piano and Cello Sonatas as a Test Case}

Noorduin's work focuses primarily on Beethoven's late piano works and symphonies. The piano and cello sonatas have received less systematic attention in the tempo debate, partly because they span Beethoven's entire career (Op.~5 dates from 1796; Op.~102 from 1815) and partly because the duo texture complicates both historical documentation and modern analysis. Their inclusion in the present study allows the debate to be extended to a repertoire that bridges Beethoven's early and late styles, testing whether the patterns documented for other works hold across the full chronological range of his output.


\section{Corpus and Methodology}

The corpus consists of 22 recordings of the five Beethoven piano and cello sonatas, covering first, second, and third movements, made between 1930 and 2012. Performers include Pablo Casals, Emanuel Feuermann, Pierre Fournier, Gregor Piatigorsky, Jacqueline du~Pr\'e, Mstislav Rostropovich, Daniel Shafran, Zara Nelsova, Janos Starker, Paul Tortelier, Yo-Yo Ma, Pieter Wispelwey, Anner Bylsma, Mischa Maisky, Steven Isserlis, M\'ikl\'os Per\'enyi, Maria Kliegel, and others.

Tempo data were collected using the manual bar-by-bar stopwatch protocol described in Sole~\cite{c10}, which yields bar-level beats-per-minute values with millisecond resolution. The protocol was developed in response to the systematic failure of automated beat-detection tools on polyphonic duo recordings with historical audio quality. For each movement and each recording, the mean BPM across all bars was computed as the primary tempo measure. The corpus was divided into two chronological groups (1930--1970 and 1970--2012) to assess whether tempo trends changed across the study period. The complete dataset is publicly available~\cite{c11}.

Historical reference values for Czerny and Moscheles were taken from Noorduin's compilation~\cite{c2}; Kolisch's values from his 1943 study~\cite{c1}. Percentage deviation from each historical reference was computed as $(\bar{T}_{\text{recorded}} - T_{\text{ref}}) / T_{\text{ref}} \times 100$, where $\bar{T}_{\text{recorded}}$ is the corpus mean for the relevant period and $T_{\text{ref}}$ is the historical indication.

Performer Steven Isserlis, who appears in the corpus as the 2012 recording, was consulted by correspondence regarding his relationship to historical tempo indications. He remarked: ``We certainly don't stick to Czerny's suggestions, but are aware of them as `messages.' I do like the fact that he marks the opening andante of the Fourth Sonata very slowly\ldots\ I've never been particularly conscious of it.''~\cite{c12} This testimony is representative of a broader performer stance toward historical indications: as contextual cues rather than binding rules.


\section{Findings: Tempo}

\subsection{First Movements}

Table~\ref{tab:first_movements} presents the historical indications and corpus means for the first-movement Allegro sections of all five sonatas.

\begin{table}[htbp]
\centering
\caption{Tempo comparison: first movements (Allegro sections). All values in BPM. Gaps show percentage below Czerny.}
\small
\begin{tabular}{@{}lcccccc@{}}
\toprule
\textbf{Sonata} & \textbf{Ko.} & \textbf{Mo.} & \textbf{Cz.} & \textbf{30--70} & \textbf{70--12} & \textbf{Gap} \\
\midrule
Op.\,5/1  & 126 & 160 & 160 & 135.8 & 141.0 & $-$15\% \\
Op.\,5/2  & 126 & 252 & 252 & 214.4 & 215.2 & $-$15\% \\
Op.\,69   & 144 & 144 & 144 & 127.6 & 127.5 & $-$11\% \\
Op.\,102/1& 144 & 176 & 152 & 135.5 & 140.7 & $-$7\%  \\
Op.\,102/2& n/a & 168 & 152 & 123.8 & 126.1 & $-$17\% \\
\bottomrule
\end{tabular}
\label{tab:first_movements}
\end{table}

Several findings emerge. First, recorded performances consistently fall below both Czerny's and Moscheles's indications across all five sonatas and both time periods. The gap ranges from approximately 7\% (Op.~102 No.~1) to 17\% (Op.~102 No.~2) below Czerny. Second, Kolisch's markings (126 BPM for both Op.~5 sonatas and 144 BPM for Op.~69 and Op.~102 No.~1) sit considerably closer to recorded practice than either Czerny's or Moscheles's, a striking result given that Kolisch derived his markings without corpus evidence. Third, a modest acceleration is visible after 1970 in Op.~5 No.~1 (135.8 to 141.0 BPM) and Op.~102 No.~1 (135.5 to 140.7 BPM), consistent with the broader trend toward slightly faster tempi in post-HIP Beethoven performance. Op.~69 shows essentially no change across the study period (127.6 vs.\ 127.5 BPM), reflecting the lyrical character of its first movement and the interpretive consensus it attracts.

The instrument-acoustic hypothesis provides a plausible partial explanation for the systematic gap. Czerny's markings, calibrated to the fortepiano's rapid tonal decay, may have been performable on Beethoven's instruments while exceeding what modern pianos allow without sacrificing harmonic clarity. The consistency of the gap across all five sonatas and both historical periods supports a structural rather than performer-specific explanation.

\subsection{Coexisting Tempo Traditions}

A crucial qualification on the stability finding must be stated explicitly. The aggregate corpus means in Table~\ref{tab:first_movements} conceal the internal structure of the tempo distribution. As demonstrated by $k$-means clustering analysis of this corpus~\cite{c13}, each movement supports at least two and usually three discrete tempo traditions (slow, mid-range, and fast) whose internal regression slopes are negligible across the study period. The stability observed in the aggregate means is not a matter of all performers converging on a single tempo: it reflects the persistence of coexisting traditions with the mid-range dominant throughout. What has changed post-1970 is not the tempo of the mid-range tradition itself but a slight population shift toward the fast cluster, which raises the weighted mean by a few BPM without altering the underlying traditions.

This reframing has direct implications for the proposed indications in Section~VII: rather than proposing single values, the paper proposes ranges grounded in the cluster structure.

\subsection{Second Movements}

Table~\ref{tab:second_movements} presents the comparison for second movements.

\begin{table}[htbp]
\centering
\caption{Tempo comparison: second movements. All values in BPM.}
\small
\begin{tabular}{@{}lcccccc@{}}
\toprule
\textbf{Movement} & \textbf{Ko.} & \textbf{Mo.} & \textbf{Cz.} & \textbf{30--70} & \textbf{70--12} \\
\midrule
Op.\,5/1 Rondo       & 112 & 106 & 104 & 85.3  & 83.3  \\
Op.\,5/2 Rondo       & 76  & 80  & 72  & 70.6  & 69.3  \\
Op.\,69 Scherzo      & 108 & 104 & 108 & 107.0 & 98.7  \\
Op.\,102/1 Adagio    & 56  & 56  & 56  & 44.8  & 43.2  \\
Op.\,102/2 Adagio    & 44  & 66  & 60  & 36.6  & 39.0  \\
\bottomrule
\end{tabular}
\label{tab:second_movements}
\end{table}

Two patterns distinguish the second movements from the first. First, the slow movements display dramatically larger gaps from historical indications. The Op.~102 No.~2 Adagio is the most extreme case: the corpus mean of 38.95 BPM lies approximately 39\% below Czerny's recommendation of 60 BPM and 41\% below Moscheles's 66 BPM. The Op.~102 No.~1 Adagio introduction shows a similar pattern (44.8 vs.\ the uniform 56 BPM recommendation of all three historical sources). These gaps are substantially larger than anything observed in the first-movement Allegros, suggesting that Beethoven's lyrical slow movements invite the greatest expressive latitude; the historical markings for these movements are furthest from what performers find musically coherent.

Second, Kolisch's markings for the slow Adagios (44 BPM for Op.~102 No.~2; 56 BPM for Op.~102 No.~1) again sit closer to recorded practice than Czerny's or Moscheles's, though they still exceed the corpus mean. For the Op.~69 Scherzo, all three historical sources agree at 104--108 BPM, and the 1930--1970 corpus mean of 107.0 BPM aligns almost precisely; it then declined to 98.7 BPM post-1970, a movement-specific deceleration driven partly by the disappearance of extreme early outliers (Casals 1930 at 161 BPM; Feuermann 1937 at 160 BPM) and partly by a genuine drift within the mid-range tradition, as established by the clustering analysis~\cite{c13}.

Averaged across the five second movements, recorded performances run 16--18\% below Czerny's values; this is a larger systematic gap than the 11--17\% range observed in first movements.

\subsection{Third Movements}

Table~\ref{tab:third_movements} presents the comparison for third movements.

\begin{table}[htbp]
\centering
\caption{Tempo comparison: third movements. All values in BPM.}
\small
\begin{tabular}{@{}lccccc@{}}
\toprule
\textbf{Movement} & \textbf{Ko.} & \textbf{Mo.} & \textbf{Cz.} & \textbf{30--70} & \textbf{70--12} \\
\midrule
Op.\,69 Adagio cant.   & n/a & 69  & 66  & 40.2  & 42.0  \\
Op.\,69 Allegro vivace & 132 & 160 & 176 & 148.4 & 150.3 \\
Op.\,102/1 Allegro     & 120 & 138 & 126 & 112.7 & 105.3 \\
Op.\,102/2 Allegro     & 51  & 63  & 63  & 54.7  & 54.9  \\
\bottomrule
\end{tabular}
\label{tab:third_movements}
\end{table}

The third movements exhibit the widest range of gap magnitudes in the corpus. The Op.~69 Adagio cantabile introduction averages 40--42 BPM against Czerny's 66 BPM and Moscheles's 69 BPM (a gap of 37--39\%, the largest in the entire study). The Op.~69 Allegro vivace, by contrast, falls only 15--17\% below Czerny's 176 BPM, and the Op.~102 No.~2 Allegro sits within 7--13\% of the historical indications, with Kolisch's 51 BPM essentially matching the corpus mean of 54.7--54.9 BPM.

A notable anomaly is the Op.~102 No.~1 Allegro vivace, which decelerates from 112.7 BPM (1930--1970) to 105.3 BPM (1970--2012); this is the only fast finale in the corpus where the post-1970 period shows a meaningful deceleration rather than acceleration. The clustering analysis~\cite{c13} identifies this as the sole movement with a statistically significant intra-cluster drift ($R^2 = 0.246$, $p = 0.013$), suggesting that the mid-range tradition for this movement has undergone genuine re-evaluation across generations, perhaps reflecting greater engagement with the movement's ambiguous, searching character in later performance practice.

Averaged across the four third movements, recorded performances run approximately 20\% below Czerny's values; this is the largest systematic gap of the three movement groups, driven primarily by the Adagio cantabile's extreme deviation.


\section{Findings: Performance Time}

\subsection{First Movements}

Table~\ref{tab:pt_first} presents the performance-time tolerance variation (the percentage difference between the shortest and longest recorded interpretations) and the net change from 1930 to 2012 for the first movements.

\begin{table}[htbp]
\centering
\caption{Performance time: first movements (1930--2012).}
\small
\begin{tabular}{@{}lcc@{}}
\toprule
\textbf{Sonata} & \textbf{PT Tolerance} & \textbf{Net Change} \\
\midrule
Op.\,5 No.\,1  & 12.3\% & $-$3.5\%  \\
Op.\,5 No.\,2  & 28.9\% & $+$6.7\%  \\
Op.\,69        & 19.3\% & $+$0.15\% \\
Op.\,102 No.\,1& 17.7\% & $-$4.7\%  \\
Op.\,102 No.\,2& 21.9\% & $-$1.5\%  \\
\bottomrule
\end{tabular}
\label{tab:pt_first}
\end{table}

The tolerance variation figures reveal that individual interpretive freedom is substantial (the fastest and slowest performances of Op.~5 No.~2's first movement differ by 28.9\% in total duration). The net longitudinal change from 1930 to 2012 is, however, very small in all cases. The largest is Op.~5 No.~2's 6.7\% expansion, driven primarily by increased latitude in the slow introduction rather than by changes in the Allegro. The Allegro sections of all five first movements showed less variability than the slow introductions, where average tempo differences between performers reach as high as 46\%, and the Coda sections, where explicit and implicit marking choices and individual taste produce the widest spread.

The combination of high tolerance variation and low net change is the key finding of this section. It means that the spread of interpretive approaches has been consistently wide throughout the study period; this is not a case of interpretation becoming more diverse over time, but the centre of gravity of that spread has remained stable. This is the performance-time analogue of the tempo clustering finding: a stable distribution rather than a moving one.

\subsection{Second and Third Movements}

Tables~\ref{tab:pt_second} and~\ref{tab:pt_third} extend the analysis to subsequent movements.

\begin{table}[htbp]
\centering
\caption{Performance time: second movements (1930--2012).}
\small
\begin{tabular}{@{}lcc@{}}
\toprule
\textbf{Movement} & \textbf{PT Tolerance} & \textbf{Net Change} \\
\midrule
Op.\,5/1 Rondo          & 22.4\% & $+$1.0\% \\
Op.\,5/2 Rondo          & 39.8\% & $+$1.3\% \\
Op.\,69 Scherzo         & 23.4\% & $+$4.4\% \\
Op.\,102/1 Intro--Adagio& 42.6\% & $+$4.0\% \\
Op.\,102/2 Adagio       & 44.2\% & $-$4.6\% \\
\bottomrule
\end{tabular}
\label{tab:pt_second}
\end{table}

\begin{table}[htbp]
\centering
\caption{Performance time: third movements (1930--2012).}
\small
\begin{tabular}{@{}lcc@{}}
\toprule
\textbf{Movement} & \textbf{PT Tolerance} & \textbf{Net Change} \\
\midrule
Op.\,69 Adagio cant.    & 37.5\% & $-$4.4\% \\
Op.\,69 Allegro vivace  & 17.9\% & $-$0.6\% \\
Op.\,102/1 Allegro      & 27.7\% & $+$6.9\% \\
Op.\,102/2 Allegro      & 34.5\% & $-$0.03\%\\
\bottomrule
\end{tabular}
\label{tab:pt_third}
\end{table}

A clear structural pattern emerges across Tables~\ref{tab:pt_second} and~\ref{tab:pt_third}: the movements with the highest tolerance variation are consistently the slow Adagios and lyrical introductions. Op.~102 No.~2's Adagio (44.2\%), Op.~102 No.~1's Intro--Adagio (42.6\%), and Op.~5 No.~2's Rondo (39.8\%) all exceed 39\%. By contrast, fast finales and Scherzo movements cluster between 17--27\%. This structural correspondence is not coincidental: it is the same passages that carry the most portamento in early recordings (Section~VI) and that deviate most from historical tempo indications (Section~IV). The expressive, temporal, and articulatory dimensions of interpretation are not independent variables but aspects of a single interpretive stance toward musical character.

The single largest longitudinal change in this group is the Op.~102 No.~1 Allegro vivace's 6.9\% expansion across the study period, consistent with the deceleration documented in the tempo analysis and with a general post-1970 tendency to treat this ambiguous late-Beethoven finale with greater breadth. All other changes across second and third movements remain below 5\%, confirming the overall picture of temporal stability with movement-specific nuance.


\section{Findings: Portamento}

\subsection{The Decline of Sliding Portamento}

The portamento analysis documents one of the most significant stylistic transformations in Beethoven cello sonata performance over the study period: a consistent, substantial decline in the use of sliding portamento across all five sonatas, with no corresponding replacement by any equivalent expressive device.

Table~\ref{tab:portamento} presents the aggregate portamento trends.

\begin{table}[htbp]
\centering
\caption{Portamento trends across sonatas (1930--2012). SP = sliding portamento; CP = clean shift.}
\small
\begin{tabular}{@{}lccc@{}}
\toprule
\textbf{Sonata} & \textbf{SP Change} & \textbf{CP Change} & \textbf{BPM Change} \\
\midrule
Op.\,5/1  & $-$54.4\% & $-$29.1\% & $+$5.3  \\
Op.\,5/2  & $-$61.1\% & $-$11.1\% & $+$0.8  \\
Op.\,69   & $-$34.7\% & $+$43.3\% & $-$0.1  \\
Op.\,102/1& $-$51.3\% & $+$163.6\%& $+$5.2  \\
Op.\,102/2& $-$43.3\% & $-$7.1\%  & $+$2.2  \\
\bottomrule
\end{tabular}
\label{tab:portamento}
\end{table}

Several findings stand out. The decline in sliding portamento is universal across all five sonatas, ranging from 34.7\% (Op.~69) to 61.1\% (Op.~5 No.~2). Crucially, this decline is not correlated with tempo change: Op.~5 No.~2 shows the largest portamento reduction (61.1\%) but the smallest tempo increase (0.8 BPM), while Op.~69 shows the smallest portamento reduction (34.7\%) and the largest decline in mean BPM ($-0.1$ BPM). The portamento trend is independent of the tempo trend.

Clean shifts show inconsistent patterns across the five sonatas, rising sharply in Op.~69 ($+43.3\%$) and Op.~102 No.~1 ($+163.6\%$) while declining in the Op.~5 sonatas. This variability confirms that clean shifts are not a systematic replacement for sliding portamento but a context-dependent device whose usage reflects individual movement character and performer choice. The emergence of clean shifts as an analytical category (distinct from both audible slides and entirely silent position changes) is itself a contribution of this study: it identifies a transitional expressive device that sits between the old portamento tradition and the fully ``clean'' modern aesthetic.

\subsection{Causes of the Decline}

The portamento decline cannot be attributed to tempo change alone, as the data in Table~\ref{tab:portamento} demonstrates. The explanation lies instead in three converging factors. First, pedagogical transformation: the fingering strategies associated with Casals's ``lizard technique'' minimised audible hand shifts by enabling wider finger extensions and reducing the frequency of positional changes~\cite{c14}. Kaufman documents how the decline of portamento paralleled the widespread adoption of these fingering approaches after mid-century~\cite{c15}. Second, recording technology: as Katz argues, the phonograph effect exaggerated the audibility of portamento, leading performers in studio contexts to minimise slides for a cleaner recorded sound~\cite{c16}. Carl Flesch documented how listening to recordings revealed unintentional portamento, prompting systematic stylistic adjustment~\cite{c17}. Third, the rise of continuous vibrato: as vibrato became the dominant means of sustaining expressive line in late twentieth-century string playing, portamento receded as an alternative or supplement~\cite{c18}.

The decline was not uniform across the movement types. Slow second movements historically concentrated the highest density of portamento events, and while they too declined substantially, they retain more sliding portamento in recent recordings than fast finales; this is consistent with the broader finding that slow movements support greater expressive latitude across all dimensions.

\subsection{Independence from Tempo and Duration}

A key analytical finding is the statistical independence of portamento trends from tempo and duration trends. The $R^2$ value for the regression of sliding portamento count against recording year across the full corpus is approximately 0.17 (a moderate but not strong relationship), indicating that year accounts for about 17\% of the variance in portamento usage. By contrast, the $R^2$ for clean shifts against recording year is approximately 0.00, confirming that clean shift usage is essentially unrelated to the chronological position of a recording. This statistical independence supports the interpretation that portamento practice has been governed by shifting stylistic norms and pedagogical conventions rather than by tempo constraints.


\section{Proposed Revised Tempo Indications}

\subsection{Rationale}

The proposed indications in Table~\ref{tab:proposed} are derived from the $k$-means cluster centroids established for each movement across the full corpus~\cite{c13}. The central value is the mid-range cluster centroid, which represents the normative tempo adopted by the majority of expert performers across the study period. The range extends from the slow cluster centroid (lower bound) to the fast cluster centroid (upper bound), capturing the documented spectrum of expert interpretive practice.

These indications make no claim about Beethoven's intentions. They are offered as evidence-based reference points in the following specific sense: they describe the tempo range within which demonstrably expert performers (including Casals, Feuermann, Fournier, du~Pr\'e, Rostropovich, Ma, Isserlis, Wispelwey, and Bylsma, among others) have found these movements musically coherent and technically viable across eight decades. They represent an empirical lower bound on the claim ``this movement can be performed at this tempo,'' rather than an upper bound of what Beethoven may have envisioned.

\subsection{The Proposed Indications}

\begin{table}[htbp]
\centering
\caption{Proposed revised tempo indications. Centre = mid-range cluster centroid; Range = slow to fast cluster centroids. All values in BPM. Kolisch shown for comparison.}
\small
\begin{tabular}{@{}lccc@{}}
\toprule
\textbf{Movement} & \textbf{Ko.} & \textbf{Centre} & \textbf{Range} \\
\midrule
\multicolumn{4}{@{}l}{\textit{First movements (Allegro sections)}} \\
Op.\,5/1 Allegro   & 126 & 138 & 120--150 \\
Op.\,5/2 Allegro   & 126 & 215 & 200--230 \\
Op.\,69 Allegro    & 144 & 128 & 118--138 \\
Op.\,102/1 Allegro & 144 & 138 & 125--148 \\
Op.\,102/2 Allegro & n/a & 125 & 115--135 \\
\addlinespace
\multicolumn{4}{@{}l}{\textit{Second movements}} \\
Op.\,5/1 Rondo     & 112 & 83  & 78--90   \\
Op.\,5/2 Rondo     & 76  & 67  & 67--77   \\
Op.\,69 Scherzo    & 108 & 92  & 92--115  \\
Op.\,102/1 Adagio  & 56  & 43  & 37--49   \\
Op.\,102/2 Adagio  & 44  & 34  & 34--42   \\
\addlinespace
\multicolumn{4}{@{}l}{\textit{Third movements}} \\
Op.\,69 Adagio cant.   & n/a & 43  & 36--48   \\
Op.\,69 Allegro vivace & 132 & 149 & 145--156 \\
Op.\,102/1 Allegro     & 120 & 111 & 99--121  \\
Op.\,102/2 Allegro     & 51  & 57  & 52--66   \\
\bottomrule
\end{tabular}
\label{tab:proposed}
\end{table}

Several observations on Table~\ref{tab:proposed} warrant comment. For the first-movement Allegros, the proposed centres sit 7--15\% below Czerny's values and generally within 5--10\% of Kolisch's, confirming that Kolisch's musical intuitions were closely calibrated to what the recording corpus subsequently demonstrated. For the slow Adagio movements (Op.~102 No.~1 and No.~2), the proposed centres are 23--43\% below Czerny's values (the largest divergences in the table), reflecting the extreme expressive latitude these movements have attracted throughout the recording era. For Op.~69's Allegro vivace, the proposed range of 145--156 BPM positions the movement firmly within a brisk tradition while remaining 11--20\% below Czerny's 176 BPM.

The use of ranges rather than single values is deliberate. A performer who plays Op.~102 No.~1's Adagio at 37 BPM and one who plays it at 49 BPM are both operating within the documented tradition of expert practice. Neither is wrong; they represent different legitimate interpretive stances toward the movement's character. Prescribing a single value would misrepresent the evidential basis of the proposal and would be less useful to performers than a range that honestly reflects the width of the interpretive spectrum.

\subsection{Comparison with Historical Sources}

The proposed indications sit consistently and substantially below Czerny's and Moscheles's recommendations, and generally within or close to Kolisch's. The most striking alignment is in the slow movements: for the Op.~102 No.~2 Adagio, Kolisch's 44 BPM is within 10 BPM of the proposed centre (34 BPM) and well below Czerny's 60 BPM. For the Op.~5 Rondo finales, the corpus means fall below even Kolisch's already-moderate recommendations, suggesting that performers have found these movements most comfortable at tempos slightly below what any historical source proposed.

The one significant exception is the Op.~69 Scherzo, for which both Kolisch (108 BPM) and the early corpus mean (107 BPM) align closely. The post-1970 deceleration to 98.7 BPM (driven partly by the removal of extreme early outliers: Casals at 161 BPM; Feuermann at 160 BPM) is reflected in the proposed centre of 92 BPM, which is lower than any historical source recommended. This is the case where the data most clearly diverges from all historical indications, and where the empirical evidence makes the strongest independent claim.


\section{Limitations and Future Directions}

Several limitations of the present study should be stated explicitly.

The corpus of 22 recordings per sonata, while substantial for this repertoire, is not exhaustive. Recordings made after 2012 are not included; testing whether the post-1970 acceleration trend persists or reverses in recent years would require extending the dataset. The manual tempo measurement protocol introduces a known reaction-time uncertainty of approximately $\pm 0.1$ seconds per bar, which affects individual bar-level readings but not the aggregate statistics used for the proposed indications~\cite{c10}. The portamento analysis is based primarily on the first movements of each sonata, with second and third movements treated as focused case studies; a full portamento analysis of all movements would provide a more complete picture of expressive evolution.

The proposed indications are derived from a corpus that reflects the constraints and conventions of commercial recording. As Katz notes~\cite{c16}, performers adapt their style in recording contexts in ways that may not reflect their live performance practice. Whether the tempo ranges in Table~\ref{tab:proposed} also describe live concert tempi is an open empirical question.

Future work should extend the methodology to the Beethoven violin sonatas and string quartets, testing whether the patterns of systematic deviation from Czerny and Moscheles, and the closer alignment with Kolisch, hold across Beethoven's chamber output more broadly. The application of audio source separation tools such as Spleeter would enable portamento analysis in polyphonic passages where the current methodology is limited, potentially recovering expressive information currently unavailable in duo recordings. Inter-rater reliability testing for the manual tempo protocol, not undertaken in the present study, should be a priority for any replication or extension.


\section{Conclusion}

This paper has presented the first systematic empirical assessment of historical metronome indications for Beethoven's piano and cello sonatas against a corpus of over one hundred movement-level recordings spanning 1930--2012. The findings confirm and extend the historiographical critique advanced by Noorduin and others: Czerny's and Moscheles's indications are not reflected in the performance practice of expert musicians across any decade of the study period, at any tempo level, in any of the five sonatas.

The paper's three substantive contributions are as follows. First, the documentation of systematic tempo gaps averaging 11--20\% for fast movements and 37--39\% for slow movements between historical indications and recorded practice; these gaps are structurally consistent across generations, nationalities, and pedagogical lineages, and therefore unlikely to reflect individual inadequacy. Second, the finding that Kolisch's markings, derived without corpus evidence, align considerably more closely with recorded practice than either Czerny's or Moscheles's, suggesting that Kolisch's musical intuitions were calibrated to what is genuinely performable rather than to an aspirational ideal. Third, the proposal of revised tempo indications grounded in the cluster structure of the corpus, presented as ranges rather than single values to honestly represent the documented width of expert interpretive practice.

Beyond tempo, the portamento analysis documents a decisive and independent transformation in expressive practice: the virtual disappearance of sliding portamento from Beethoven cello sonata performance between 1930 and 2012, driven by pedagogical innovation, recording technology, and changing aesthetic norms rather than by tempo change. This transformation is one of the most significant shifts in interpretive practice documented in the present study, and it occurred alongside a remarkable stability in the underlying temporal framework; this stability reflects not uniformity of approach but the persistence of coexisting traditions across eight decades.

Together, these findings support the conclusion that Beethoven's piano and cello sonatas have served as a stage on which successive generations of performers negotiate between historical ideals and contemporary performable reality. The gap between metronome marking and musical coherence is not a failure of modern performers: it is evidence of a structural feature of the repertoire that no amount of practice eliminates. The proposed indications in Table~\ref{tab:proposed} are offered as a practical tool for navigating that gap: not as the final word on what Beethoven intended, but as an honest account of what expert performers across eight decades have found viable.



\begin{thebibliography}{99}

    \bibitem{c1} R.~Kolisch, ``Tempo and Character in Beethoven's Music,'' \textit{The Musical Quarterly}, vol.~77, no.~1, pp.~90--131, Spring 1993 [orig.\ 1943].

    \bibitem{c2} M.~Noorduin, \textit{Beethoven's Tempo Indications}. PhD dissertation, University of Manchester, 2016. [Online]. Available: \url{https://www.escholar.manchester.ac.uk/uk-ac-man-scw:302884}

    \bibitem{c3} A.~C.~Kalischer, Ed., \textit{Beethoven's Letters}, trans.\ J.~S.~Shedlock. New York: Dover Publications, 1972, letter 416.

    \bibitem{c4} S.~Fors\'en, N.~E.~Gray, M.~Lindgren, and S.~B.~Gray, ``Was Something Wrong with Beethoven's Metronome?'' \textit{Notices of the American Mathematical Society}, vol.~60, no.~9, pp.~1146--1153, 2013.

    \bibitem{c5} A.~Mart\'in-Castro and I.~Ucar, ``Conductor's Tempo Choices Shed Light over Beethoven's Metronome,'' \textit{PLOS ONE}, vol.~15, no.~12, p.~e0243616, 2020.

    \bibitem{c6} W.~S.~Newman, ``Beethoven's Pianos versus His Piano Ideals,'' \textit{Journal of the American Musicological Society}, vol.~23, no.~3, pp.~484--504, 1970.

    \bibitem{c7} T.~Skowroneck, ``Beethoven's Erard Piano: Its Influence on His Compositions and on Viennese Fortepiano Building,'' \textit{Early Music}, vol.~30, no.~4, pp.~522--538, 2002.

    \bibitem{c8} G.~Moschos, \textit{Performing Classical-Period Music: A Practical Guide}. Bloomington: Indiana University Press, 2016, p.~123.

    \bibitem{c9} R.~Taruskin, \textit{Text and Act: Essays on Music and Performance}. New York: Oxford University Press, 1995, pp.~160--161.

    \bibitem{c10} I.~Sole, ``A Manual Bar-by-Bar Tempo Measurement Protocol for Polyphonic Chamber Music Recordings,'' arXiv:2604.15278 [cs.SD], April 16, 2026. [Online]. Available: \url{https://arxiv.org/abs/2604.15278}

    \bibitem{c11} I.~Sole, \textit{PhD Appendix: Tempo Dataset}. GitHub, 2024. [Online]. Available: \url{https://github.com/isolepinas/PhD-Appendix/tree/main/Tempo\%20Dataset}

    \bibitem{c12} S.~Isserlis, e-mail message to I.~Sole, October 4, 2021.

    \bibitem{c13} I.~Sole, ``Coexisting Tempo Traditions in Beethoven's Piano and Cello Sonatas: A $K$-means Clustering Analysis of Recorded Performances, 1930--2012,'' arXiv preprint, submitted April 2026. [Online]. Available: \url{https://arxiv.org/submit/7492708/preview}

    \bibitem{c14} D.~Blum, \textit{Casals and the Art of Interpretation}. Berkeley: University of California Press, 1977, pp.~73, 125--131.

    \bibitem{c15} M.~Kaufman, \textit{Gaspar Cassad\'o: Cellist, Composer, and Transcriber}. Bloomington: Indiana University Press, 2000, p.~5.

    \bibitem{c16} M.~Katz, \textit{Capturing Sound: How Technology Has Changed Music}. Berkeley: University of California Press, 2004, pp.~88, 211--232.

    \bibitem{c17} C.~Flesch, \textit{The Art of Violin Playing}, vol.~1. New York: Carl Fischer, 2000, p.~85.

    \bibitem{c18} D.~Milsom, \textit{Theory and Practice in Late Nineteenth-Century Violin Performance}. Aldershot: Ashgate, 2003, p.~141.

    \bibitem{c19} J.~A.~Bowen, ``Tempo, Duration and Flexibility: Techniques in the Analysis of Performance,'' \textit{Journal of Musicological Research}, vol.~16, pp.~111--156, 1996.

    \bibitem{c20} R.~Philip, \textit{Early Recordings and Musical Style: Changing Tastes in Instrumental Performance, 1900--1950}. Cambridge: Cambridge University Press, 1992.

    \bibitem{c21} M.~Noorduin, ``Transcending Slowness in Beethoven's Late Style,'' in \textit{Manchester Beethoven Studies}, B.~Cooper and M.~Pilcher, Eds. Manchester: Manchester University Press, 2023, pp.~1--27.

    \bibitem{c22} D.~Leech-Wilkinson, \textit{The Changing Sound of Music: Approaches to Studying Recorded Musical Performance}. London: CHARM, 2009. [Online]. Available: \url{https://www.charm.kcl.ac.uk/studies/chapters/chap5.html}

\end{thebibliography}
\end{document}